\documentclass{acm_proc_article-sp}
\usepackage{graphicx}
\usepackage{amssymb}

\begin{document}

\title{Active Flows in Diagnostic of Troubleshooting on Backbone Links\titlenote{The research is supported partially by grant N06-07-89074 of RFBR}}

\numberofauthors{5} %  in this sample file, there are a *total*
\author{
% 1st. author
\alignauthor
Andrei M. Sukhov\titlenote{corresponding author}\\
       \affaddr{Samara State Aerospace University}\\
       \affaddr{Moskovskoe sh., 34}\\
       \affaddr{Samara, 443086, Russia}\\
       \email{amskh@yandex.ru}
% 2nd. author
\alignauthor
Dmitry I. Sidelnikov\\
       \affaddr{Institute of Organic Chemistry of RAS}\\
       \affaddr{Leninsky pros., 47}\\
       \affaddr{Moscow, 119991, Russia }\\
       \email{sid@free.net}
\alignauthor
Alexey Galtsev \\
       \affaddr{Samara State Aerospace University}\\
       \affaddr{Moskovskoe sh., 34}\\
       \affaddr{Samara, 443086, Russia}\\
       \email{galaleksey@gmail.com}
\and
\alignauthor      
% 3rd. author
Alexey P. Platonov\\
       \affaddr{Russian Institute for Public Networks} \\
       \affaddr{Kurchatova sq. 1}\\
       \affaddr{Moscow, 123182, Russia}\\
       \email{plat@ripn.net}
% 4th. author
\alignauthor 
Mikhail V. Strizhov\\
         \affaddr{Samara State Aerospace University}\\
       \affaddr{Moskovskoe sh., 34}\\
       \affaddr{Samara, 443086, Russia}\\
       \email{strizhov@ip4tv.ru}
}

\date{10 November 2009}

\maketitle

\begin{abstract}
This paper aims to identify the operational region of a link in terms of its utilization and alert operators at the point where the link becomes overloaded and requires a capacity upgrade. The number of active flows is considered the real network state and is proposed to use a proxy for utilization. The Gaussian approximation gives the expression for the confidence interval on an operational region. The easy rule has been formulated to display the network defects by means of measurements of router loading and number of active flows. Mean flow performance is considered as the basic universal index characterized quality of network services provided to single user.
\end{abstract}

\category{C.2.3}{Computer-communication networks}{Network Operations}[network monitoring]
\category{C.4}{Performance of systems}{Measurement techniques}

\keywords{Network States on Flow Level, Mean Flow Performance, Test for Network Quality} % NOT required for Proceedings

\section{Introduction}

Typically, the following four values are used for the estimation of the network quality:

\begin{itemize}
	\item 
Link or router loading (or available bandwidth for end-to-end connection) 
	\item 
Round trip time 
	\item 
Packet loss rate
	\item 
IP packet delay variation (network jitter)
\end{itemize}

The available bandwidth, round trip time, packet loss rate and IP packet delay variation describe the quality
of connectivity between two remote points or end-to-end connection. The link utilization is
applied to the monitoring of a single hop between two routers~\cite{ftd,ptz}.

Network operators need to know when their backbone or peering links must be upgraded.
Boundary values of network parameters may serve as an indicator e.g. as the current
values of the network parameters reach a defined limit, the links have to be upgraded.
The problem with this method is that there is no standardized set of network parameters
to monitor. Each provider has its own set of technical specifications aimed at avoiding
overload. Big providers, like Sprint \cite{ftd}, rely on the results of their own research. Usually,
network operators monitor peak and average link utilization levels and upgrade their links
when the utilization level is in the range 30\%-60\%.

The main focus of this paper is to use flow-based analysis~\cite{bmr} to monitor the backbone
link and identify when an upgrade is needed. Previous work by Chuck Fraleigh et al \cite{ftd}
addressed a similar provisioning problem to reduce the {\em per packet} end-to-end delay. Dina
Papagiannaki et al \cite{ptz}, at Infocom 03, introduced a methodology on the basis of SNMP
statistics to predict when and where link additions/upgrades should take place in an IP
backbone network.

For resource management in IP networks, several recent
proposals~\cite{cfg} advocate for a flow oriented architecture. A flow
is classically identified by the usual 5tuple composed of the
source and destination addresses together with the source
and destination port numbers and protocol type. The basic
principles of flow-aware networking rely on the fact that flows
are the elementary entities associated with user transactions.
In order to provide end users with an acceptable quality, it
is essential to share the bandwidth of the network by taking
into account flows~\cite{bbp,jen,kmo,ouro}. To efficiently implement
flow aware resource management techniques, however, it is
essential to estimate the number of flows simultaneously active
on a link in an operational network~\cite{apg,deri}.

Traffic accounting mechanisms based on flows should be considered as passive measurement
mechanisms. Information is gathered by flows are useful for many purposes:
\begin{itemize}
	\item 
Understanding the behaviour of existing networks
	\item 
Planning for network development and expansion
	\item 
Quantifying network performance
	\item 
Verifying the quality of network service
	\item
Attribution of network usage to users	
\end{itemize}
Unfortunately, today there is no common view on how to estimate the connection quality, and how to
find bottlenecks in networks.

Barakat et al \cite{bti} propose a model that relies on flow-level information to compute the
total (aggregate) rate of data observed on an IP backbone link. For modelling purposes,
the traffic is viewed as the superposition (i.e. multiplexing) of a large number of flows that
arrive at random times and that stay active for random periods.

Our paper presents a technique for estimating the network behaviour based on the utilization
curve which is representing correlation between link utilization and the number of
active flows in it. We implicitly argue that the number of active flows may be
considered as the {\em real} network state~\cite{apg,ygd} and is consequently, a better indication of utilization or
desired operating point.

Our objective is to gather knowledge and plot the curve for network quality and related
terms for utilization such as: length of operational region, mean flow performance, confidence interval, points of overload, etc. Traffic measurement and analysis is extended to consider the network quality for a large network or for a high-speed backbone~\cite{hlf,nss}. This will help discover possible bottlenecks on external links of enterprize network or backbones~\cite{ftd}. We locate the threshold point at which the addition of new flows does not
increase the link utilization.

In order to prove our hypothesis we took measurements from the border gateway routers of Russian Internet Service Provider FREEnet. FREEnet (The Network For Research, Education and Engineering), an academic and research network, was founded on July 20, 1991, by the N.D.Zelinsky Institute of Organic Chemistry (IOC RAS) at the Center of
Computer Assistance for Chemical Research (CACR). It assists and fulfills the networking needs of the research institutes of Russian Academy of Sciences, universities, colleges, and other research and academic institutions of Russia. FREEnet provides IPv4/IPv6 connectivity, worldwide multicast IP connectivity, DNS services, stratum
1 time service, mail relaying, collocation and hosting services, assorted information services. 

Total capacity of upstream links is presently 2.1 Gbps, 1.3 Gbps links interconnect FREEnet with its peers (excluding FREEnet members). There are 6 regional branches that operate independently but they are cooperating as peers in accordance with FREEnet Charter principles. Tens of thousands researchers from hundreds institutions countrywide enjoy FREEnet services.

Earlier, in order to prove our hypothesis we took measurements from the border gateway routers
of Russian ISP ``SamaraTelecom'' (ST) and from HEAnet - Ireland\rq s National Research and
Education Network.

In the course of experiment average flow performance in investigated networks considerably differs. So flow performance in provincial Russian networks more then one order less similar indicator HEAnet. These results are completely not surprising; essentially they reflect quality of network services. At once there was an idea to apply the model found us for the comparative analysis of large networks or date-centers. Thus it is possible to use unique parameter - mean flow performance.

This paper describes an apllication of simple flow-based model \cite{apg,bti} to identify the quality 
of the connection
and the instance when a backbone upgrade is required. We present our findings under the
following headings:
\begin{itemize}
	\item
Section~\ref{sb} - using queueing theory for flow-based analysis of a backbone
	\item
Section~\ref{s2} discusses three states of network on flow level
	\item
Section~\ref{s4} - a test for network quality
	\item
Section~\ref{s5} - results from experiments conducted in the real ISP\rq s
	\item
Section~\ref{s6} describes techniques of comparison of a communication quality in the large networks.
\end{itemize}

\section{The Flow Based Model}
\label{sb}

In this paper we present traffic as a stationary process, using the results from the papers of
Barakat et al~\cite{bti} and Ben Fredj et al~\cite{bbp}. They proposed a traffic model for uncongested
backbone links that is simple enough to be used in network operations and engineering. The
model of Barakat et al relies on Poisson shot-noise. With only 3 parameters ($\lambda$, arrival rate
of flows, ${\Bbb E}[S_n]$, average size of a flow, and ${\Bbb E}[S^2_n/D_n]$, average value for the 
ratio of the square of a flow size and its duration), the model provides approximations for the average of
the total rate (the throughput) on a backbone link and for its variations at short timescales.
The model is designed to be general so that it can be easily used without any constraints
from the definition of flows, or on the application or the transport protocol. In summary,
this model allows us to completely characterize the data rate on a backbone link based on
the following inputs:
\begin{itemize}
	\item 
Session arrivals in any period where the traffic intensity is approximately constant
are accurately modelled by a homogeneous Poisson process of finite rate $\lambda$. The
measurements of Barakat et al~\cite{bti} showed that the arrival rate $\lambda$ remains pretty constant
for at least a 30 minute interval. In general, this assumption can be relaxed to
more general processes such as MAPs (Markov Arrival Processes)~\cite{bfb}, or non homogeneous
Poisson processes, but we will keep working with it for simplicity of the
analysis.
	\item 
The distribution of flow sizes ${S_n}$ and flow durations ${D_n}$. In this paper we denote
$T_n$ as the arrival time of the $n$-th flow, $S_n$ as its size (e.g., in bits), and $D_n$ as its
duration (e.g., in seconds). Sequences ${S_n}$ and ${D_n}$ also form independently of
each other and are identically distributed sequences.

	\item 
The flow rate function (shot) is $X_n(t)$. A flow is called active at time $t$ when 
$T_n \leq t \leq T_n +D_n$. Define as $X_n(t-T_n)$ the rate of the $n$-th flow at time 
$t$ (e.g., in bits/s),
with $X_n(t - T_n)$ equal to zero for $t < T_n$ and for $t > T_n + D_n$.
\end{itemize}

Define $B(t)$ as the total rate of data (e.g., in bits/s) on the modeled link at time $t$. It is
determined by adding the rates of the different flows. We can then write
\begin{equation}
	B(t) = \sum_{n \in {\Bbb Z}} X_n(t - T_n)
\label{nag}
\end{equation}

The process from Eq.~(\ref{nag})  can describe the number of active flows $N$ found at time $t$ in an
$M/G/\infty$ queue~\cite{kl}, if $X_n(t - T_n) = 1$ at $t \in [T_n, T_n + D_n]$.

The model presented by Barakat et al~\cite{bti} can compute the average and the variation of
traffic on the backbone. In summary:
\begin{itemize}
	\item 
The average total rate of the traffic is given by the two parameters $\lambda$ and ${\Bbb E}[S_n]$:
\begin{equation}
{\Bbb E}[B(t)] = \lambda {\Bbb E}[S_n] 
\label{mB}
\end{equation}
	\item 
The variance of the total rate ${\Bbb V}[B(t)]$ (i.e., burstiness of the traffic) is given by the two
parameters $\lambda$ and ${\Bbb E}[S^2_n/D_n]$:
\begin{equation}
{\Bbb V}[B(t)] = \lambda {\Bbb E}[S^2_n/D_n]
\label{VB}
\end{equation}
\end{itemize}

It should be mentioned that Eq.~(\ref{mB}) is true only for the ideal case of a backbone link of
unrestricted capacity, that can be applied to underloaded links. The main drawback of the
ratio (\ref{mB}) is its lack of definite usage limits, due to the fact that variables 
$\lambda, {\Bbb E}[S_n]$ describing
the system are in no way connected with its current state. The average flow size ${\Bbb E}[S_n]$ does
not depend on a specific system, it is an universal value determined by the current distribution
of file sizes found in the Internet.

The arrival rate of flows $\lambda$ describes the user\rq~s behaviour and doesn\rq~t depend on the network
state and utilization. The cumulative number of flows that arrive at a link will remain
linear even if the network has problems and doesn\rq~t satisfy all the incoming demands.

In order to describe the real network state with arbitrary load we should use Little\rq~s law:
\begin{equation}
N = \lambda {\Bbb E}[D_n] 
\label{mN}
\end{equation}

Here ${\Bbb E}[D_n]$ is the mean duration of flow and $N$ is the mean number of active flows.
Formula~(\ref{mN}) is true \cite{kl}  for any flow duration and thus for an arbitrary flow size distribution
and rate limit. This formula describes the network state more precisely than Eq.~(\ref{mB}) as
the average number of active flows on the bandwidth unit increases with the utilization. In
other words, the average duration of flow ${\Bbb E}[D_n]$ enables us to judge the real network state
in contrast to its average value ${\Bbb E}[S_n]$.

\section{Network States on Flow Level}
\label{s2}

\begin{figure}
\centering
\includegraphics[width=7cm]{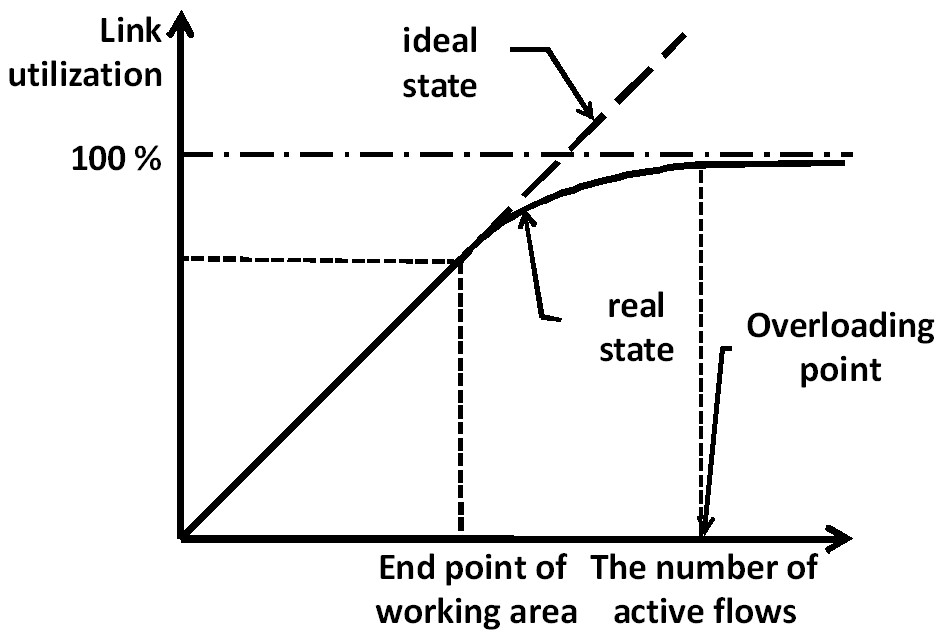}
\caption{Link utilization vs the number of active flows}
\label{f1}
\end{figure}

In order to analyze the connection quality at the backbone area or the link to the provider we are going to investigate a graphical dependence between the link utilization $U$ and the number of active flows $N$ in it~\cite{apg}. These variables are easy measurable quantities in despite of average values ${\Bbb E}[D_n]$ and ${\Bbb E}[S_n]$. The separate network state is pictured by single point on coordinate plane with axes $N$ and $U$. The curve depicting average values has been shown in Fig.~\ref{f1}.

On the curve shown in Fig.~\ref{f1}, three parts can be identified, corresponding to the different network states. The first part of the curve describes the network state close to the ideal. If the investigated link has unrestricted capacity there should be a stable linear relationship between number of lows $N$ and link utilization $U$. The curve describing the network behavior beyond a certain point will be convex. The linear part of the curve corresponding to the ideal network behavior is defined as the operational region. The operational region ends at the threshold point which should be found experimentally. The dislocation of this point depends on many factors, such as, transport layer protocol, network topology, the amount of buffering at the link, etc.

The second part of the curve corresponds to the moderately loaded network, when the diversion from
the ideal network state becomes obvious. There is an increase in the average duration of
a flow compared to the working area, and therefore, a larger number of active flows on the
bandwidth unit characteristic of this network state.

The third part of the curve corresponds to the totally disabled network with considerable
packet loss evident. We propose some simple preliminary models for an overloaded
link, accounting for user impatience and reattempt behaviour. In a real network, if demand
exceeds capacity, the number of flows in progress does not increase indefinitely. As perflow
throughput decreases, some flows or sessions will be interrupted, due either to user
impatience or to aborts by TCP or higher layer protocols.

In the end of this section  estimation of confidence interval is given for the operational 
region of our curve. Since the total rate is the result of multiplexing of $N(t)$ flows
of independent rates, the Central Limit Theorem~\cite{kl}  tells us that the distribution of $B(t)$
tends to Gaussian at high loads, which is typical of backbone links. As it is mentioned in Section~\ref{sb}, the variance of the
total rate requires two parameters: the arrival rate of flows $\lambda$ and the expectation of the
ratio between the square of the size of a flow duration ${\Bbb E}[S^2_n/D_n]$  (see Eq.~\ref{VB}). It tells us
that the total rate should lie between ${\Bbb E}[B]- A(\varepsilon)\sqrt{{\Bbb V} (B)}$
and ${\Bbb E}[B]+ A(\varepsilon)\sqrt{{\Bbb V} (B)}$
in order to
provide a required quality of service. When we talked about the required quality of service
we implied the accordance of network behavior to Eq.~(\ref{mB}), here $A(\varepsilon)$ is the $\varepsilon$-quantile of the centered
and normalized total rate $B(t)$.

Taking into account Eqs.~(\ref{mB}-\ref{mN}) and theorems about average values the confidence interval of the bandwidth $B$ 
on a operational region of our curve is
\begin{equation}
	B= b(N\pm \alpha A(\varepsilon)\sqrt{N}),
	\label{e1}
\end{equation}
here $b=E[S_n/D_n]$ is the average 
flow performance which characterizesis the speed of user communication, coefficient $\alpha$ could be found experimentally.

\section{Testbed Setup}
\label{s4}

In order to prove our hypothesis we took measurements on border gateway routers from FREEnet,
HEAnet - Ireland\rq s National Research and Education Network, and also from the Russian
ISP ``SamaraTelecom'' (ST). All networks have several internal and external links. ST\rq s
basic load lies on one channel to the Internet, whereas HEAnet and FREEnet relies on a number
of connections. Measurements from Gigabit links were taken for FREEnet and 155Mbps, 
622Mbps are for HEAnet and ST. The utilization of these links varies widely from 5\% to 60\%
with a clearly identifiable busy period.

A passive monitoring system based on Cisco\rq s NetFlow~\cite{cisco} technology was used to
collect link utilization values and active flow numbers in real-time. In Moscow and Samara we measured
on a Cisco 7206 router with NetFlow switched on. At HEAnet a Cisco 12008 was utilized.
A detailed description of Cisco NetFlow can be found in the Cisco documentation~\cite{cisco}.

This is achieved using the following commands on the Cisco 7206:
\begin{itemize}
	\item 
{\sffamily sh ip cache flow} - gives information about the number of active and inactive flows,
about the parameters of the flows in the real time.
	\item
{\sffamily show interface summary} - gives information about the current link
utilization.
\end{itemize}

On a GSR Router these commands look like:
\begin{itemize}
	\item 
{\sffamily enable}
\item
 {\sffamily attach slot-number}
\item
{\sffamily show ip cache flow}
\end{itemize}

The FREEnet data was obtained using scripts running every 30 minutes from middle of January to the end of March 2008. The data sets from two routers of FREEnet have been collected for further analysis. The full loading of first router varies in limits of hundreds megabits per second ({\em 100-220 Mbps\/}) and tens megabits for second router (see Fig.~\ref{rs1}). During the tests we fixed information about any network events that could influence on connection quality. The ST data were recorded at 30-minute intervals, twenty-four hours a day for a week to discover network behavior with different loading levels. The HEAnet data was obtained using scripts running every 5 minutes for a period of 72 hours. It is quite easy to write a script, which will collect the data from the router to the management server.

\section{Experimental Validation and Diagnostic of Troubleshooting}
\label{s5}

The network state at every instant describes by point on two-dimensional plot where abscise shows the number of active flow $N$ and ordinate shows router loading $B$. Basic tasks of experimental validation of our model consist in
\begin{itemize}
	\item 
construction of curve of average values and its comparison with theoretical prediction showed on Fig.~\ref{f1}
	\item 
calculation of variance for flow performances and verification of parabolic form of confidence interval from Eq.~(\ref{e1})
	\item 
examination on normal distribution for flow performances
\end{itemize}

In order to do experimental test of our model the data set should be divided into several intervals depending on number of active flows. Inside each interval the average values and their variance were calculated for flow performances as well as for other parameters characterizing network states. 

\begin{table}
\centering
\caption{Parameters for active flows of FREEnet, Data Set 1, 2008}
\begin{tabular}{|c|c|c|c|c|c|} \hline
\textsl{n} & $N$ & $B$,  & $\sigma (B)$, & $b$,  & $\sigma (b)$, \\ 
& & \textit{Mbps} & \textit{Mbps} & \textit{bps} & \textit{bps} \\ \hline
1& 17489 & 113.1 & 23.1 & 6784 & 1386 \\ \hline
2 & 23260 & 126.0 & 21.4 & 5682 & 965 \\ \hline
3 & 27007 & 152.0 & 39.2 & 5628 & 1452 \\ \hline
4 & 34902 & 156.7 & 26.9 & 4990 & 770 \\ \hline
5 & 45104 & 163.9 & 33.9 & 3634 & 752 \\ \hline
6 & 55019 & 176.3 & 33.2 & 3205 & 604 \\ \hline
7 & 64778 & 215.4 & 42.2 & 3325 & 652 \\ \hline
\end{tabular}
\label{t1}
\end{table}

The earlier tests have been conducted on the boundary router of the ST and HEAnet, and they didn't allow to verify the model with high precision. The new data from FREEnet contains thousands of points describing network states.

FREEnet Data Set 1 has been divided into seven intervals according to the number of active flows (15000 - 20000, 20000 - 25000, 25000 - 30000, 30000 - 40000, 50000 - 60000, $\geq$60000). Inside each interval basic parameters characterizing active flows have been calculated and the result is represented in the Table~\ref{t1}. Here $N$ is the average number of active flows for the interval denoted by $n$. $B$ describes the average router loading in \textit{Megabit per second (Mbps)\/}, and $b$ is average flow performance measured in \textit{bit per second (bps)\/}. $\sigma (b), \sigma (B)$ are the standard deviation for flow performances $b$ and router loading $B$ correspondingly.

\begin{figure}
\centering
\includegraphics[width=7cm]{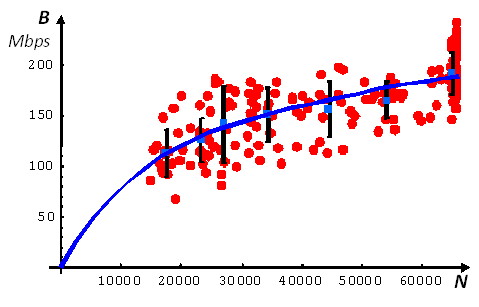}
\caption{FREEnet router loading vs the number of active flows, DataSet1}
\label{f2}
\end{figure}

The results of the measurement for FREEnet Data Set 1  are pictured on Fig.~\ref{f2}. Basic curve is constructed as the line of average values, it describes network states on flow level. The error bar restricts the confidential interval for network states. Comparison with theoretical prediction from Section~\ref{s2} leads to the conclusion that the area of network exploitation lies inside the operational region.

\begin{figure}
\centering
\includegraphics[width=7cm]{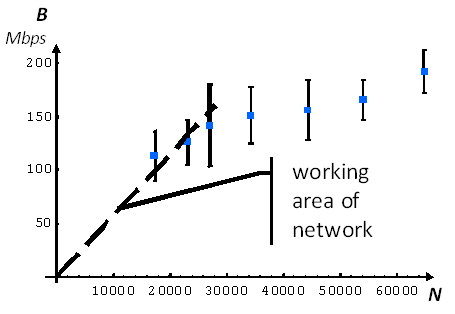}
\caption{The operational region of network, DataSet1}
\label{f3}
\end{figure}

In order to restrict the operational region the straight portion of curve from Fig.~\ref{f2} should be marked as it is shown on Fig.~\ref{f3}. The slope angle of this straight line is found as the average flow performance $b$. Only three initial points from investigated data set may be placed in the limits of operational region. The angle of inclination gives the average flow performance equal to 5700 \textit{bps} for FREEnet. If the number of flows exceeds 30000 then network gets moderately loading that leads to a reduction of the flow performance. The router loading does not increase uniformly with the number of requests, and the connection quality becomes almost twice as bad (see Table~\ref{t1}).

\begin{table}
\centering
\caption{Parameters for active flows of FREEnet, Data Set 2, 2008}
\begin{tabular}{|c|c|c|c|c|c|} \hline
\textsl{n} & $N$ & $B$,  & $\sigma (B)$, & $b$,  & $\sigma (b)$, \\ 
& & \textit{Mbps} & \textit{Mbps} & \textit{bps} & \textit{bps} \\ \hline
1 & 5446 & 15.42 & 2.25 & 2843 & 413 \\ \hline
2 & 6531 & 17.11 & 2.45 & 2364 & 377 \\ \hline
3 & 7508 & 17.74 & 2.35 & 2364 & 313 \\ \hline
4 & 8370 & 18.92 & 2.24 & 2261 & 268 \\ \hline
5 & 9443 & 20.67 & 3.81 & 2190 & 404 \\ \hline
6 & 15495 & 28.05 & 5.40 & 1811 & 349 \\ \hline
\end{tabular}
\label{t2}
\end{table}

\begin{figure}
\centering
\includegraphics[width=7cm]{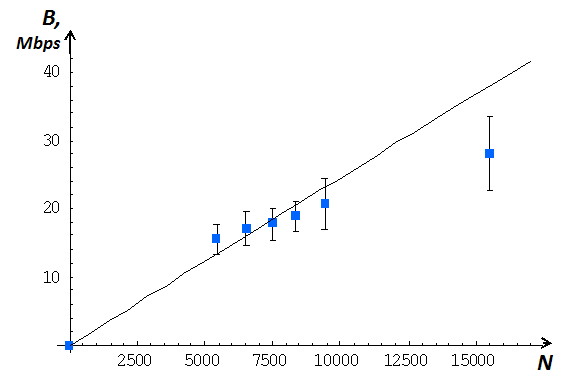}
\caption{The operational region of network, DataSet2}
\label{fd3A}
\end{figure}

In Table~\ref{t2} the Data Set 2 from the second router of FREEnet with lower loading is presented. Investigated region divides into six intervals according to the number of active flows (5000 - 6000, 6000 - 7000, 7000 - 8000, 9000 - 10000, $\geq$10000). Operational region for second router of FREEnet is restricted by 10 000 active flows as it is shown in Fig.~\ref{fd3A}. Only last intervals should be excluded from straight portion of utilization curve.

\begin{figure}
\centering
\includegraphics[width=7cm]{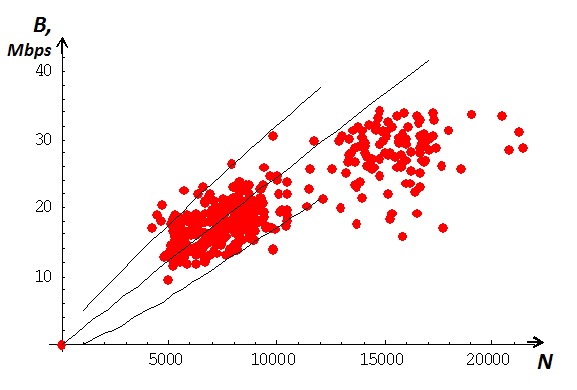}
\caption{Confidence interval for operation region, DataSet2 from FREEnet}
\label{rs1}
\end{figure}

Fig.~\ref{rs1} illustrates the network states and the form of confidence interval for operation region with normal quantile function $A(\varepsilon)=0.05$.

A correlation coefficient indicates the strength and direction of a linear relationship between two random variables. In order to verify the parabolic form of confidence interval the correlation coefficient between variables $\sigma_i(B)$ and $\sqrt{N_i}$ should be calculated for both DataSet of FREEnet (see Tables~\ref{t1} and \ref{t2}). Comparing second and forth columns of mentioned Tables the values of correlation coefficient are equal 0.70 for DataSet1 and 0.93 for DataSet2. These magnitudes allow us to say about high correlation between theoretical model and its experimental examination.

A significant question concerns the numerical value for the the numerical coefficient $\alpha_1, \alpha_2$
from Eq.~(\ref{e1}). The function $A(\varepsilon)$ can be computed using the Gaussian approximation,
which gives for example $A(0.05) = 1.96$. Our data from Tables~\ref{t1},~\ref{t2} allow calculating their magnitudes:
\begin{equation}
	\alpha_1 \approx 13, \alpha_2 \approx 4.5
	\label{e2}
\end{equation}

\begin{table}
\centering
\caption{Statistical tests, Data Set 1, 2008}
\begin{tabular}{|c|c|c|c|} \hline
\textsl{n} &  $\chi^2$ for &  Gaussian & Correlation \\ 
&  $\alpha=0.95$ & test &  coefficient \\ \hline
1 &  not enough data &  &  \\ 
2 &  not enough data &  &  \\ 
3 &  not enough data &  &  \\ 
4 &  3.49 (9.49) & $+$ & 0.70 \\ 
5 &  not enough data &  &  \\ 
6 &  3.45 (7.81) & $+$ &  \\ 
7 &  0.50 (9.49)& $+$ &  \\ \hline
\end{tabular}
\label{t3}
\end{table}

Significant assumption underlies a theoretical model that distribution of flow performances $b$ may be considered as Gaussian distribution. The number of testing network states inside many intervals from Tables~\ref{t1},~\ref{t2} allow to check a given set for similarity to the normal distribution. Here we use the Pearson $\chi^2$ test and the results of this test could be found in Tables~\ref{t3} and \ref{t4}. Column 2 shows the value of $\chi^2$  for $\alpha=0.95$,  in round brackets it is shown Table values for $\chi^2$. It should be noted that all investigated intervals with sufficient number of states discover the normal type of distribution.

\begin{table}
\centering
\caption{Statistical tests, Data Set 2, 2008}
\begin{tabular}{|c|c|c|c|} \hline
\textsl{n} &  $\chi^2$ for &  Gaussian & Correlation \\ 
&  $\alpha=0.95$ & test &  coefficient \\ \hline
1 &  9.15 (12.6) & $+$ &  \\ 
2 &  10.0 (11.1) & $+$ &  \\ 
3 &  3.24 (14.1) &  $+$& 0.93 \\ 
4 &  10.0 (14.1) & $+$ &  \\ 
5 &  not enough data &  &  \\ 
6 &  1.94 (14.1)& $+$ &  \\ \hline
\end{tabular}
\label{t4}
\end{table}

In conclusion of this Section it should be noted that expression~(\ref{e1}) allows formulating the easy rule how to display the network defects. If two consistent measurements running every 5(30) minutes give the deviation of real network states $B_i,N_i$ from the confidence interval with $A(0.05)$ then network problem has been detected. Confidence interval is described by flow performance $b$ and the values $\alpha,\sigma(B)$ which may be found only as result of data processing. Unfortunately, the corresponding software is not completed yet.

\begin{figure}
\centering
\includegraphics[width=7.5cm]{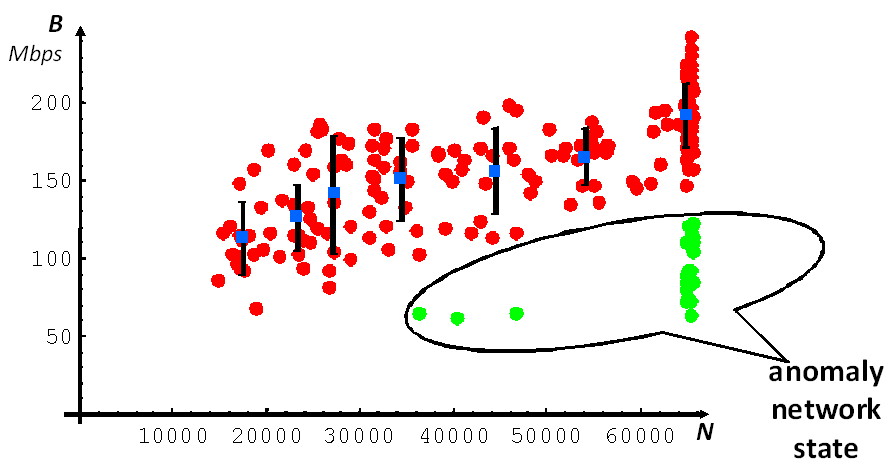}
\caption{Detection of anomaly network state}
\label{f4}
\end{figure}

This rule received an apt illustration during FREEnet network testing. A network incident has been detected: a wide links to large FTP server has been temporally turned down. These anomaly network states $B_i, N_i$ departure the confidence interval corresponding to standard behavior of investigated network and form separate cluster as it has shown on Fig.~\ref{f4}.

So our model receives the experimental confirmation and the diagnostic method based on introduction of confidence interval may be applied to network monitoring.

\section{Parameters of Comparative Studies}
\label{s6}

During experiments it has been noticed that the average flow performances in investigated networks are considerable different. 
 
For different Russian networks this parameter varies from \textit{hundreds bps\/} to \textit{tens Kbps\/}, for regional network it exceeds 2\textit{Kbps\/} very seldom. For comparison, the mean flow performance in HEAnet differs on almost one order and equals 15000 \textit{bps\/} that allows them to provide high speed video applications through public network. In other words the investigated Russian network needs expanding bandwidth of trunk links, especially those ones that connect remote regions. It is problem number one for Russian Networks for Science and Education.

The idea evolved to apply our model for comparative analysis of large networks or data-center connectivity. The equation (\ref{e1}) allows us to compare the length of operation regions, mean flow performances, the width of confidence intervals, etc. 

It should be noted that the curve of mean values from Fig.~\ref{f2} describes the behavior of investigated network as a whole. Mean flow performance $b$ allows to judge about the quality of network services provided to single user. The calculation of active flow number has a significant feature; the flow is considered active for a long time after transmission of the last packet. Therefore the real speed of network services exceeds the value $b$ almost by one order.

From the end user's point of view the mean flow performance is the basic universal index. Principally, this value should be considered as the basic index characterizing the quality of network services. 

There is a dependence of access to high-speed Internet services and mean flow performance. For example, streaming video began possible to look at our university as soon as the given indicator has exceeded value in 10 \textit{Kbps\/}. It is very much a rough estimate, additional researches are necessary to find boundary values of mean flow performance, necessary for introduction of this or that high-speed service. This problem in networks of cellular operators where standards of data transmission GPRS, EDGE and even 3G are not capable to give yet high speeds to an unlimited circle of clients is especially actual.

\section{Conclusion}
\label{s7}

In this paper some methods are described that allow us to evaluate connection quality in large networks based on flow technology, and give an indication as to when capacity needs to be increased. At the moment we are working on developing utilities, which will make it possible to construct the dependence of link loading on the number of active flows automatically, and calculate the length of the operation region as well as coefficients for confidence interval.

The special attention has been given how by means of the constructed model to make a technique of comparison of the big networks. It has been established that as object for comparison the unique parameter - mean flow performance can act. The high-grade comparative analysis should contain the conformity table between mean flow performance and possibility of start high-speed Internet services that demands the further experiments.

This paper has demonstrated that it is possible to easily determine the confidence interval, operation region and the overload point of a network connection, utilizing low cost commodity hardware and simple software. This allows us to identify the anomaly network states and moment when a backbone upgrade is required. Further experiments are necessary in order to develop software utilities for this purpose. Thus, providing analytical generalizations, we established common terminology for processes, taking place in networks.

\end{document}